\documentclass[12pt,preprint]{aastex}

\slugcomment{Accepted by ApJ Letters April 9, 2009.}

\shorttitle{Multiple Populations in 47~Tuc} 
\shortauthors{Anderson et al.} 
\usepackage{ulem}

\begin{document}

\title{MIXED POPULATIONS IN GLOBULAR CLUSTERS:\ \\
          {\it ET TU}, 47 TUC?\footnote{
                               Based on observations with  the
                               NASA/ESA {\it Hubble Space Telescope},
                               obtained at  the Space Telescope Science
                               Institute,  which is operated by AURA, Inc.,
                               under NASA contract NAS 5-26555.}}

\author{J.\ Anderson\altaffilmark{2},
          G.\  Piotto\altaffilmark{3},
          I.\ R.\ King\altaffilmark{3},
          L.\ R.\ Bedin\altaffilmark{2}, and
          P.\ Guhathakurta\altaffilmark{4}}

\altaffiltext{2}{Space Telescope Science Institute,
                  3800 San Martin Drive, Baltimore,
                  MD 21218; [jayander,bedin]@stsci.edu}

\altaffiltext{3}{Astronomy   Dept., Padova Univ.,
                  Vic. Osservatorio 2, 35122, PD, Italy;
                  piotto@pd.astro.it}

\altaffiltext{4}{Astronomy Dept., Univ. of Washington,
                  Box 351580, Seattle, WA 98195-1580, USA;
                  king@astro.washington.edu}

\altaffiltext{5}{UCO/Lick Observatory,
                  University of California Santa Cruz,
                  1156 High St., Santa Cruz, CA 95064;
                  raja@ucolick.org}

\begin{abstract}
We exploit the large number of archival {\it HST} images of 47 Tuc
to examine its subgiant branch (SGB) and main sequence (MS) for signs 
of multiple populations.  In the cluster core, we find that the cluster's 
SGB exhibits a clear spread in luminosity, with at least two distinct 
components:\ a brighter one with a spread that is real but not bimodal, 
and a second one about 0.05 mag fainter, containing about 10\% of the stars.
In a less crowded field 6\arcmin\  from the center, we find that the 
MS is broadened much more than can be accounted for by photometric errors,
and that this broadening increases at fainter magnitudes.
\end{abstract}


\section{Introduction}
\label{S.intro}

For half a century the standard doctrine that each globular cluster (GC)
consists of stars born at the same time out of the same material has
contributed enormously to our understanding of stellar evolution.  In
recent years, however, discoveries made largely by our group have
contradicted this paradigm, and the study of GC populations has moved in
a new direction.  It had already been noted that the CMDs of M54 
(Sarajedini \& Layden 1995) and $\omega$~Cen (Lee et al.\ 1999 and 
Pancino et al.\ 2000) showed evidence of multiple red-giant branches,
but the new era began with the discovery (Bedin et al.\ 2004) that 
the main sequence of $\omega$ Centauri is split into two branches. 
Spectroscopy showed (Piotto et al.\ 2005) that the metallicities made 
sense only if the bluer branch had a strong helium enhancement.

Next, discoveries based on photometry of unparalleled accuracy, on 
{\it HST}/ACS images, have shown that NGC 2808 has three parallel 
main sequences (MSs) (Piotto et al.\ 2007), and NGC 1851 has a split 
subgiant branch (Milone et al.\ 2008a).  More recently, we have found 
split sub-giant branches (SGBs) in NGC 6388, NGC 6656, and other 
Galactic GCs (see Piotto 2009 for an update), and in many clusters 
of the Large Magellanic Cloud (Milone et al.\ 2008b).

We note in passing that Galactic GCs have also shown abundance
peculiarities (see Gratton et al.\ 2004 for the most recent review) 
that also suggest multiplicities of population, but here we will 
be discussing only photometric peculiarities.

One striking characteristic common to all the clusters that we have
mentioned is that they are among the most luminous GCs in the Galaxy.
The most massive cluster that has not yet been investigated with 
high-precision photometry is 47 Tucanae.  Fortunately, 47 Tuc has been
used for {\it HST} calibration, and a wealth of archival data is
available.  In this paper we will examine two fields in 47 Tuc.  In the
central field, where the star numbers are greatest, we will examine the
subgiant branch, and in an outer field, where the crowding of faint
stars is minimal, we will take a close look at the stars of the main
sequence.

\section{Multiple SGBs in the Cluster core} 
\label{S.SGBinn}

The core of 47 Tuc has been observed several times with {\it HST}'s
Advanced Camera for Surveys.  The archive contains a very large set of
observations taken with the HRC, but there are too few stars in the
small HRC field to get good statistics on the SGB population, so we
focus here on the WFC data.  Five data sets were used here 
(program-ID, filters, number of exposure, exposure time):\ 
(1) GO-9028  F475W 20$\times$60 s, 
(2) GO-9443  F475W  5$\times$60 s, 
(3) GO-9503  F475W  1$\times$60 s, 
(4) GO-9281  F435W  9$\times$105 s and
               F625W 20$\times$65 s, 
(5) GO-10775 F606W  4$\times$50 s and
               F814W  5$\times$50 s.

The {\tt \_flt} images from the {\it HST} pipeline were reduced 
using the program described in Anderson \& King (2006, AK06), which uses 
for each filter a spatially varying array of PSFs, plus a ``perturbation
PSF'' tailored to each exposure to compensate for the effect of focal 
variations.  The photometry was put into the ACS/WFC Vega-mag system 
according to the procedure given by Bedin et al.\ (2005), and using the 
encircled energy and zero points given by Sirianni et al.\ (2005).

Features on the subgiant branch are usually difficult to discern, since
the fiducial sequence moves rapidly in both color and magnitude.  Here,
we can take advantage of the fact that for the metallicity and age of 
47 Tuc, the SGB is nearly horizontal in F475W, so that all SGB stars 
should have the same F475W flux, within observational errors.

The first three programs that we listed give us a F475W observation 
for each star.  The next two programs each give us a useful color
measurement.  We thus have for each star three independent F475W
magnitude measurements and two independent measurements of the color, 
$(m_{\rm F435W}-m_{\rm F625W})$ and $(m_{\rm F606W}-m_{\rm F814W})$.
To ensure the best possible photometry, when we reduced each of the 
images we retained only the unsaturated stars that gave a good fit 
to the PSF (the {\tt QSEL} flag, see AK06).  This did a good job 
of preventing crowded stars with compromised photometry from entering 
our lists (particularly important in the crowded core of 47 Tuc). 
Next, we made a master list of stars from the GO-9028 data set, 
and cross-identified the photometry in all the other data sets 
with this.

We now had seven independent fluxes for each star:\ three in F475W and
one in each of F435W, F625W, F606W, and F814W.  Figure~1 shows the six
CMDs that can be constructed from the three independent $m_{\rm F475W}$\
magnitudes and the two colors.  In the top left plot, which shows the
best photometry, we arbitrarily select three groups of stars:\ those
well below the SGB (red open triangles), those in the lower half of the
main SGB (green open circles), and those in its upper half (blue
filled circles).  The black lines in the first CMD delineate the 
regions chosen.  We plot these same stars in all the other diagrams.  
In the first column, the stars are plotted according to their observed 
colors in the GO-9281 data set; in the second column we use the colors 
from GO-10775.  The three panels down each column use the F475W magnitude 
from a different one of the three independent F475W data sets, which 
were taken at different pointings, times, and orientations.  The fact 
that the red-triangle stars lie below the main SGB in all the plots means 
that they truly are fainter.  Similarly, the fact that the blue filled-circle
stars are brighter than the green open-circle ones also confirms that 
the magnitude difference is significant and shows an intrinsic vertical 
spread in the main SGB.

The histograms on the right show the distribution of the stars in F475W 
magnitude as measured in the three data sets.  It is clear that the
stars that were shown as blue filled circles are indeed systematically 
brighter than those that were shown as green open circles.  There is no 
apparent bimodality in these two groups, but our multiple independent 
observations demonstrate that the observed spread is not consistent with 
random photometric errors.  The SGB stars that were shown as red triangles
are considerably fainter ($\sim$0.05 magnitude) than the other two groups; 
they make up roughly 10\% of the SGB population.

\begin{figure}
\plotone{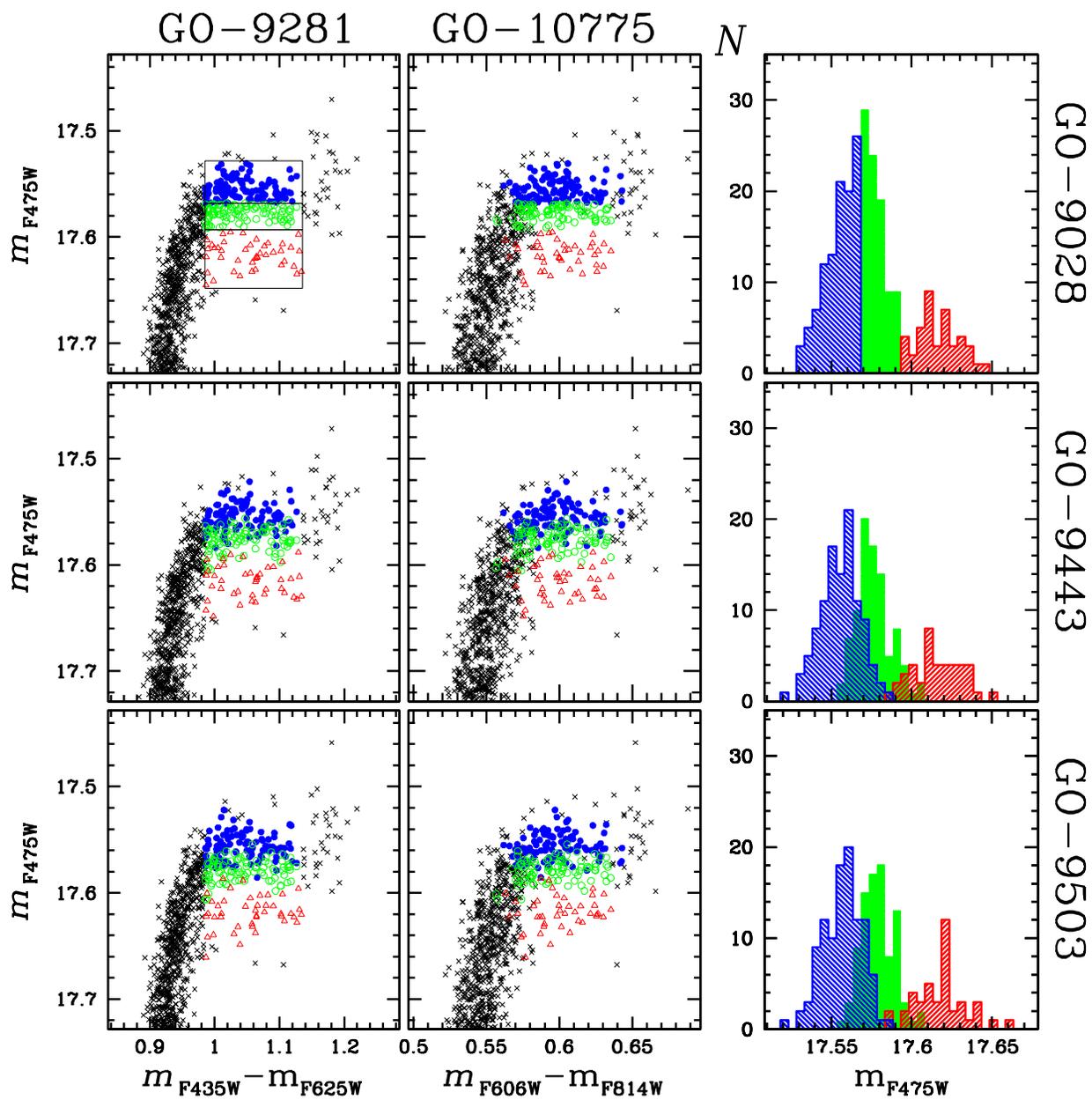}
\caption{Independent observations demonstrate that the intrinsic
           broadening of the SGB of 47 Tuc must be real (see text).
           \label{fig1}}
\end{figure}

47 Tuc has also been extensively observed in an outer field, located
6$'$ West of the cluster center.  These observations will be used in the
following section to study the MS.  We also examined the SGB there,
using the short exposures:  the SGB stars are not numerous enough for 
a detailed study, but they show strong indications of a vertical spread similar 
to what we saw in the inner field.  Studying the radial gradient of the 
SGB populations will require a larger survey than our single outlying 
ACS field can provide.

\section{Intrinsic Breadth of the MS}
\label{S.MS_width}

While the density in the outer field at 6$'$ is too low for studying the
SGB in detail, it is ideal for examining the structure of the main
sequence.  In addition to the short exposures mentioned briefly above, 
there are also more than a hundred deep exposures, taken at a wide 
variety of epochs, offsets, and orientations, for both calibration 
and scientific purposes.  In toto, there are 115 images in F606W, with
exposure times from 339 to 1400 s, from 10 different GO numbers, and 
32 in F814W, of 339 to 1500 s, from seven different GO's.

We will use this material to show that the MS of 47 Tuc has an intrinsic
width much greater than the spread that would be produced by
observational error alone; the key to this is of course establishing as
reliably as possible how small our observational errors are.  With data
sets coming from such varied circumstances, combining them to get an
accurate color and magnitude for each star is a complicated procedure,
and it would be hopelessly naive to use any sort of $\sqrt{N}$ rule;
what we do instead is to divide our data into two halves, each having
random orientations and offsets, and estimate our final errors from the
star-by-star agreement between the results from the two halves.

Our first step, however, was to remove the one systematic error that we
could correct, the small photometric error that depends on position on
the detector chips.  Using the entire outer-field data set, we derived a
residual for each star in each image, by comparing each measured F606W
magnitude with the average F606W magnitude of that star in all 115 F606W
images, and similarly for each F814W image against the average from all
32 F814W images.  We then examined the residuals for each filter as a
function of location on each chip, and constructed for each filter a
128$\times$128 array of corrections (spaced at 32-pixel intervals),
which we applied to our photometry.  These corrections were typically
0.005 mag in size; they compensate for the fact that our 9$\times$10
array of library PSFs does not allow perfectly for all PSF variations.

We made the division of our data set into halves as follows.  We first
divided the images into four sets:\ the first half of the F606W images
(V1) and the second half (V2), and the first and second halves of the
F814W images (I1 and I2).  We constructed a CMD from the V1 and I1 data,
and another independent one from the V2 and I2 data.  In each CMD we
drew in the same main-sequence ridge line (MSRL), and subtracted from 
the color of each star the color of the MSRL at its measured V magnitude, 
thus creating for each half of the data a vertical sequence with a spread 
in color.

\begin{figure}
\plotone{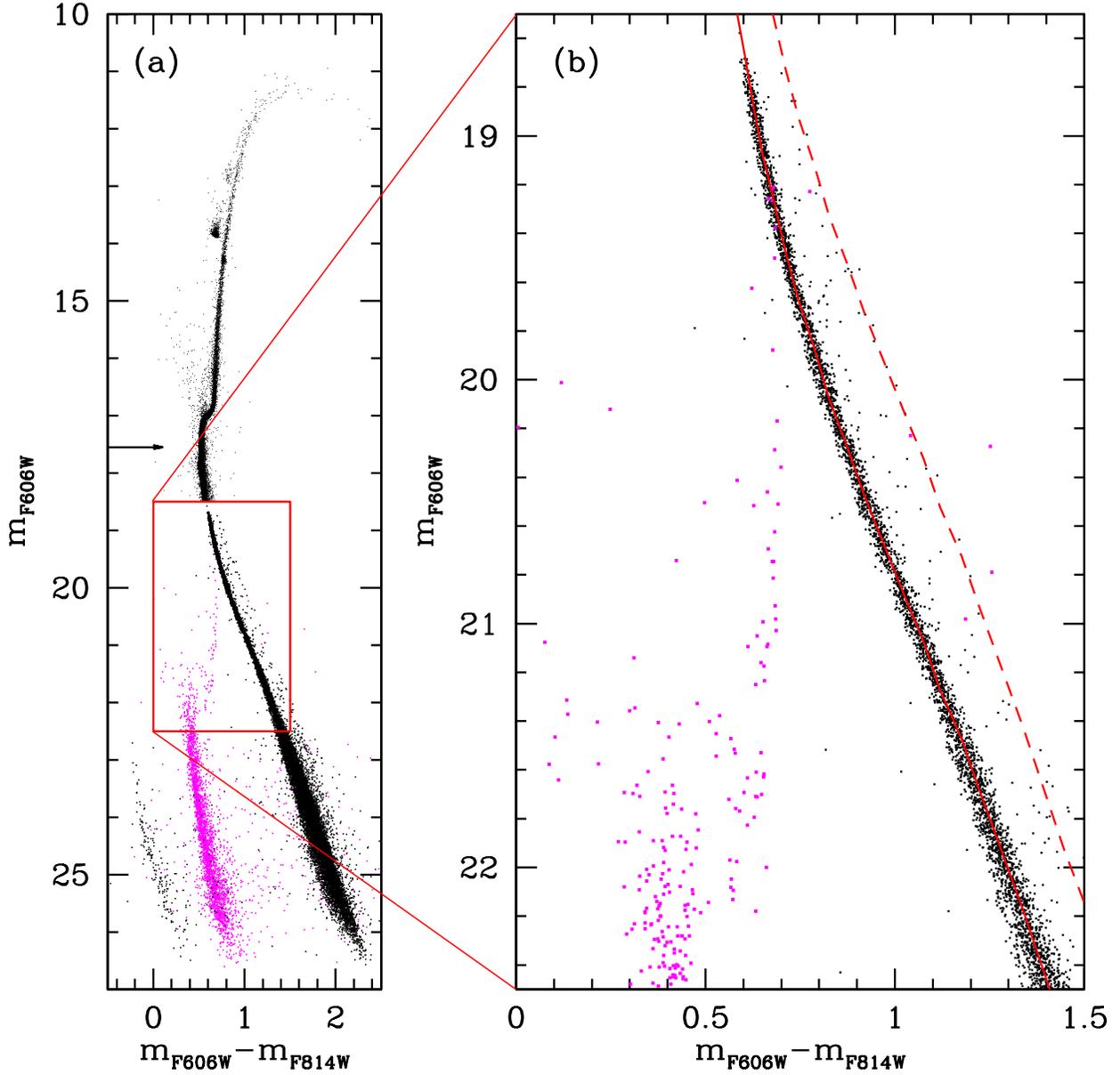}
\caption{(a) The full CMD, with proper-motion non-members in magenta.
             We mark the location of the main-sequence turnoff
             with an arrow just below $m_{\rm F606W} \sim 17.5$.
         (b) Region with high-quality deep photometry, with the
             MSRL and equal-mass-binary line drawn in.
           \label{fig2a}}
\end{figure}

Figure~\ref{fig2a} provides some orientation for our star measurements
in the outer field.  Panel (a) shows a CMD of 47 Tuc, with the faint
stars coming from the analyzed outer field, and the upper part coming
from the central field (where there are more bright stars).  Panel (b) is 
a blowup identifying the magnitude range with which the analysis in this 
section deals.  Stars in magenta are non-members, according to their 
proper motions; they are not used in what follows.  (Most of the non-members
are SMC stars in the background.  The cluster members at lower left are 
white dwarfs.) The dashed red line in panel (b) is the locus of equal-mass
MS binaries.

\begin{figure}
\plotone{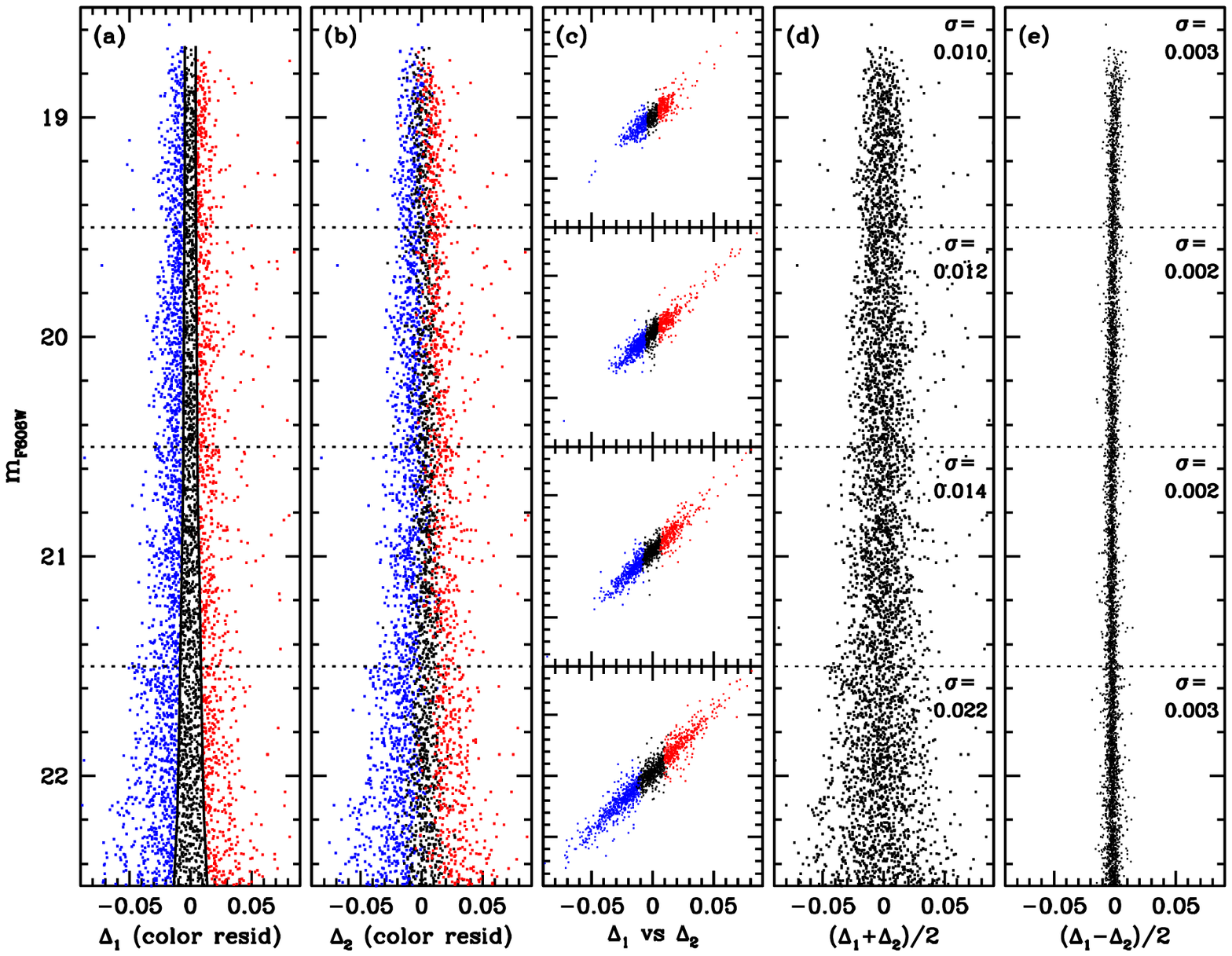}
\caption{(a) Straightened sequence for the 1st half of the data,
             divided into blue, middle, and red thirds.
         (b) Straightened sequence for the 2nd half of the data;
             the color for each star as was assigned in the
             previous panel.
         (c) Correlation of the two independent color residuals,
             for the four magnitude bins. 
         (d) Mean values of color residuals.
         (e) Estimates of color errors.
           \label{fig2b}}
\end{figure}

Figure~\ref{fig2b} shows the results of our analysis of the deep 
outer-field data.  Panel (a) shows the color residual, $\Delta_1$, 
for each the star as obtained from the first half of the data, in the sense 
$(m_{F606W} - m_{F814W})_{obs} - (m_{F606W} - m_{F814W})_{MSRL}$.  
The second color residual $\Delta_2$, in panel (b), shows the same 
quantity for the second half of the data.  These two estimates of 
the color residual of each star relative to the MSRL should be entirely 
independent of each other.  The vertical black lines in panel (a) are 
drawn in such a way that there are equal numbers of points to the 
blue, to the red, and between the two lines, and the points on 
either side are color-coded accordingly.  In panel (b) each star 
retains the color-coding that was assigned to it in panel (a); it 
is striking how well the ordering of these colors is maintained in 
this independent sample.

To emphasize the consistency of the results from the two halves of our
data, in panel (c) we plot $\Delta_1$ against $\Delta_2$.  The tight
correlation shows that each star is measured to be bluer or redder than
the MSRL by nearly the same amount in the independent halves of the
data.  This is one way of demonstrating the reality of the apparent
color width of the MS.  The spread from lower left to upper right is 
the color spread of the individual stars, while the spread in the
perpendicular direction is indicative of our measuring error.  (It is 
interesting also to note, in the upper two sections, the unilateral 
spread of binaries redward of the MS; this fails to be evident in 
the lower two sections only because of the increased spread of the 
MS stars.)

The next two panels show the same relationships in a more quantitative
way.  The quantities $\Delta_1$ and $\Delta_2$ are independent 
measurements of the color offset between each star and the MSRL.  To 
obtain the best estimate of the $\Delta$ for each star we average the 
two estimates.  To estimate the error in this average, we assume that 
$\Delta_1$ and $\Delta_2$ are Gaussian deviates with rms $\sigma$ 
and that the difference between them should be $\sim \sqrt{2}\, \sigma$, 
and that the error in the average should then be $\sigma/\sqrt{2}$, 
or $\sim |\Delta_1-\Delta_2|$/2.  In panels (d) and (e), each of the 
quoted sigmas was derived by taking half the the distance between the 
16th and 84th percentiles of the distribution in that magnitude
bin.

The error estimates are admittedly that; in the presence of small 
systematic errors of $\sim$0.005 magnitude in each exposure due to 
uncompensated-for breathing and other effects, it is difficult to be 
sure of the exact size of our errors.  Another more tediously derived 
estimate (based on a study of many different pairings of F606W and 
F814W exposures) had, for example, an increase of the error sigma from
0.0012 to 0.0028 from brightest stars to faintest.  But what is 
unmistakably true in any case is that the observed spread in MS colors
is 5 to 10 times what can be explained by measuring error.  {\it The 
main sequence of 47 Tuc has a considerable intrinsic breadth.} 
Furthermore, the increase in breadth from bright to faint is greater 
than can be explained by an increase in measurement error, and must 
therefore be real.

Finally, we note that the average MSRL residuals in panel (d) show 
no structure; like the upper branch of the split SGB, the MS shows 
no signs of a split.

\section{Discussion}

The CMDs of 47Tuc presented in this paper show that this cluster 
hosts a complex stellar population, as do all the other massive
($M>10^6m_\odot$) clusters investigated so far.  In particular: 
$i)$ The SGB is split into two SGBs, with the upper SGB showing 
in addition a clear vertical broadening.   $ii)$ The MS also shows 
an intrinsic broadening, which tends to increase as we go from 3
to 6 magnitudes below the turn-off.

The complex SGB morphology is hard to interpret without a detailed
chemical analysis of the stars in the different SGBs.  Very likely it
is the result of a combination of metallicity, age, and possibly He
variations, as in the case of $\omega$~Cen (Villanova et al.\ 2007) and
NGC 1851 (Cassisi et al.\ 2008).

A broadening of the MS at $>3$ magnitudes below the turn-off could 
in principle be due to some combination of 1) a depth effect, 
2) binaries, 3) differential reddening, 4) a spread in metallicity,
or 5) a spread in He.  A simple King model shows that the line-of-sight 
distribution of stars at this location in the cluster would imply 
a spread of less than 0.005 magnitude in the photometry, and a spread 
of 0.001 in color relative to the MSRL.  This clearly cannot explain 
the spread that we see.  As for a binary explanation, equal-mass 
binaries would have $\Delta \gtrsim 0.15$ mag in color, so the small 
symmetric spread observed would imply a conspiracy of a very high 
binary fraction and a distribution with very low secondary/primary 
mass ratios.  Besides, it seems even more unlikely that any distribution
of binaries could explain the symmetry of the color spread.  Variable 
reddening is also unlikely to explain the observed spread, since 
the total reddening is only $E(B-V)=0.04$ (Harris 1996).

We have used isochrones from the Teramo group (Pietrinferni et 
al.\ 2004), recalculated by S.\ Cassisi for different helium contents,
to estimate the Fe and He dispersion that could lead to the observed 
color spread in the MS.  For $m_{\rm F606W}$ between 20 and 21, 
we have an average $(m_{\rm F606W}-m_{\rm F814W})$ color dispersion of 
$\sim 0.013$ magnitude.  If the observed dispersion is due solely to 
a dispersion in iron (with standard He content), then it would imply
$\Delta$[Fe/H] $\simeq$ 0.1.  If He is the sole cause of the MS 
broadening, then a He dispersion $\Delta Y\sim 0.026$ would be implied.

It should come as no surprise that a cluster main sequence has an
intrinsic spread in color.  Indeed, there had to be some limit to the
homogeneity in the cloud that formed it, even under the
single-population paradigm.  It is still unclear, though, whether the
spread that we see in 47 Tuc is primordial or whether it is indicative 
of a more complex formation history.  It is important to emphasize that we
do not have such a high-precision data set for any other cluster, so it
is simply not possible to search other clusters for MS broadening at the
level of accuracy needed to find a color spread of $\sim 0.01$.
Nevertheless, the SGB spread of $>0.05$ magnitude in F475W is 
impossible to explain in terms of a simple, single stellar population.

Looking ahead, a spectroscopic study of the stars in the different 
SGB populations could resolve whether the vertical spread is due to age 
or metallicity (or both).  Also, by surveying how the fraction of stars 
in the SGB populations varies over a larger radial region, we could 
explore whether the multiple populations are better explained by a 
merger or self-enrichment scenario.  Such a survey would have to cover
much more than a single outer ACS field, and would require proper-motion
cleaning to distinguish the minority SGB populations from field and 
SMC stars.

\bigskip 
J.A.\ and I.R.K.\ acknowledge support from STScI grants GO-9444 and
GO-10101. GP acknowledges support by PRIN2007 and ASI under the
program ASI-INAF I/016/07/0.


\begin{thebibliography}{}

\bibitem[Anderson \& King (2006)]{ak06}
         Anderson, J., \& King, I.\ R. 2006, ACS ISR 2006-01

\bibitem[Bedin et al.\ (2004)]{b04}
         Bedin, L.\  R., Piotto, G., Anderson, J., Cassisi, S., King, I.\ R.,
         Momany, Y., \& Carraro, G.  2004, ApJL, 605, L125

\bibitem[Bedin et al.\ (2005)]{b05}
         Bedin, L.\  R., Cassisi, S., Castelli, F., Piotto, G., 
         Anderson, J., Salaris, M., Momany, Y., \& Pietrinferni, A.  
         2005, MNRAS, 357, 1038

\bibitem[Cassisi et al.\ (2008)]{c08}
         Cassisi, S., Salaris, M., Pietrinferni, A.,
         Piotto, G., Milone, A.\ P., Bedin, L.\  R., \& Anderson, J. 2008,
         ApJL, 672, L115

\bibitem[Gratton et al.\ (2004)]{g04}
         Gratton, R., Sneden, C., \& Carretta, E.  2004, ARA\&A, 42, 385

\bibitem[Harris et al.\ (1996)]{h96}
         Harris, W.\ E.  1996, AJ, 112, 1487 (February 2003 update).

\bibitem[Lee et al.\ (1999)]{l99}
         Lee, Y.-W., Joo, J.-M., Sohn, Y.-J., Rey, S.-C.,
         Lee, H.-C., \& Walker, A.\ R. 1999, Nature, 402, 55

\bibitem[Milone et al.\ (2008a)]{mil08a}
         Milone, A.\ P., et al.\ 2008a, ApJ, 673, 241
         % Milone, A. P.; Bedin, L. R.; Piotto, G.; Anderson, J.;
         % King, I. R.; Sarajedini, A.; Dotter, A.; Chaboyer, B.;
         % Marin-Franch, A.; Majewski, S.; Aparicio, A.;
         % Hempel, M.; Paust, N. E. Q.; Reid, I. N.;
         % Rosenberg, A.; Siegel, M.

\bibitem[Milone et al.\ (2008b)]{mil08b}
         Milone, A.\ P., Bedin, L.\ R., Piotto, G., \&
         Anderson, J. 2008b, A\&A, in press (arXiv:0810.2558)

\bibitem[Pancino et al.\ (2000)]{pan00}
         Pancino, E., Ferraro, F.\ R., Bellazzini, M.,
         Piotto, G., \& Zoccali, M. 2000, ApJ, 534, L83

\bibitem[Pietrinferni et al.\ (2004)]{p04}
         Pietrinferni, A., Cassisi, S., Salaris, M., \&
         Castelli, F., 2004, ApJ, 612, 168

\bibitem[Piotto et al.\  (2005)]{gp05}
         Piotto, G., et al. 2005, ApJ, 621, 777
         % Piotto, Giampaolo; Villanova, Sandro; Bedin, Luigi R.;
         % Gratton, Raffaele; Cassisi, Santi; Momany, Yazan;
         % Recio-Blanco, Alejandra; Lucatello, Sara;
         % Anderson, Jay; King, Ivan R.; Pietrinferni, Adriano;
         % Carraro, Giovanni

\bibitem[Piotto et al.\  (2007)]{gp07}
          Piotto, G., Bedin, L.\  R., Anderson, J., King, I.\  R., Cassisi, S.,
          Milone, A.\  P., Villanova, S., Pietrinferni, A., \& Renzini, A.
                  2007, ApJ, 661, L53

\bibitem[Piotto 2009]{gp09}
         Piotto, G.  2009, IAU Symposium No.\ 258, in press (astro-ph,
         arXiv:0902.1422)

\bibitem[Sarajedini \& Layden (1995)]{sl95}
         Sarajedini, A., \& Layden, A.\ C. 1995, AJ, 109, 1086

\bibitem[Sirianni et al.\ (2005)]{sir05}
         Sirianni, M., et al.\ 2005, PASP, 117, 1049
         % Sirianni, M.; Jee, M. J.; Benitez, N.; Blakeslee, J. P.;
         % Martel, A. R.; Meurer, G.; Clampin, M.; De Marchi, G.;
         % Ford, H. C.; Gilliland, R.; Hartig, G. F.;
         % Illingworth, G. D.; Mack, J.; McCann, W. J.

\bibitem[Villanova et al.\ (2007)]{v07}
         Villanova, S., et al. 2007, ApJ, 663, 296
         % Villanova, S.; Piotto, G.; King, I. R.; Anderson, J.;
         % Bedin, L. R.; Gratton, R. G.; Cassisi, S.; Momany, Y.;
         % Bellini, A.; Cool, A. M.; Recio-Blanco, A.; Renzini, A.

\end{thebibliography}
\end{document}